\documentclass[5p, preprint]{elsarticle} % twocolumn
\biboptions{sort&compress}
\usepackage{lineno,hyperref,isotope,mathrsfs}
\modulolinenumbers[5]
\usepackage{amsmath,bbold,amssymb,epsfig,bm,mathrsfs,feynmp,color,slashed,nicefrac,exscale,multirow,color,soul}
\usepackage[T1]{fontenc}
\usepackage[utf8]{inputenc}
\usepackage{times,txfonts}
\usepackage{isotope}
\usepackage{xcolor}
\usepackage{graphicx}
\usepackage{gensymb}
\usepackage{tikz}
\newcommand{\be}{\begin{equation}}
\newcommand{\ee}{\end{equation}}
\newcommand{\bea}{\begin{eqnarray}}
\newcommand{\eea}{\end{eqnarray}}

\definecolor{red}{rgb}{0.8,0,0}
\definecolor{violet}{rgb}{0.4,0,0.4}
\definecolor{green}{rgb}{0,0.5,0.0}
\definecolor{navy}{rgb}{0.0,0.0,0.6}
\definecolor{orange}{rgb}{0.8,0.2,0.0}
\usepackage[normalem]{ulem}  % \sout{old text} for strikeout

\begin{document}
\begin{frontmatter}

\title{Competition between delta isobars and hyperons and properties of compact stars}

\author[a]{Jia Jie Li}
\ead{jiajieli@itp.uni-frankfurt.de}
\author[b]{Armen Sedrakian}
\ead{sedrakian@fias.uni-frankfurt.de}
\author[c,d]{Fridolin Weber}
\ead{fweber@sdsu.edu}

\address[a]{ Institute for Theoretical Physics, J. W. Goethe University, D-60438 Frankfurt am Main, Germany}
\address[b]{Frankfurt Institute for Advanced Studies, D-60438 Frankfurt am Main, Germany}
\address[c]{Department of Physics, San Diego State University, 5500
Campanile Drive, San Diego, CA 92182, USA}
\address[d]{Center for Astrophysics and Space Sciences, University of California at San
Diego, La Jolla, CA 92093, USA}

\begin{abstract}
  The $\Delta$-isobar degrees of freedom are included in the covariant
  density functional (CDF) theory to study the equation of state (EoS)
  and composition of dense matter in compact stars. In addition to
  $\Delta$'s we include the full octet of baryons, which allows us to
  study the interplay between the onset of delta isobars and hyperonic
  degrees of freedom. Using both the Hartree and Hartree-Fock
  approximation we find that $\Delta$'s appear already at densities
  slightly above the saturation density of nuclear matter for a wide
  range of the meson-$\Delta$ coupling constants. This delays the
  appearance of hyperons and significantly affects the gross
  properties of compact stars.  Specifically, $\Delta$'s soften the
  EoS at low densities but stiffen it at high densities. This
  softening reduces the radius of a canonical $1.4 M_\odot$ star by up
  to 2~km for a reasonably attractive $\Delta$ potential in matter,
  while the stiffening results in larger maximum masses of compact
  stars. We conclude that the hypernuclear CDF parametrizations that
  satisfy the 2$M_\odot$ maximum mass constraint remain valid when
  $\Delta$ isobars are included, with the important consequence that
  the resulting stellar radii are shifted toward lower values, which
  is in agreement with the analysis of neutron star radii.
\end{abstract}


\end{frontmatter}
%\linenumbers

%-----------------
\section{Introduction}
\label{sec:intro}
%-----------------

Compact stars are unique laboratories for studies of dense hadronic
matter~\cite{Glenden_book,Weber_book,Lattimer2004,Weber2007,Sedrakian2007,Oertel2017}.
The hadronic core of a compact star extends from half up to a few
times the nuclear saturation density $\rho_0$. In the high-density
region of the core a number of exotic degrees of freedom are expected
to appear in addition to nucleons. Possible new constituents of
matter include hyperons~\cite{Ambartsumyan1960,Pandharipande1971,Glendenning1985,
Glendenning1991,Schaffner1994,HuberWeber1998,Massot2012,Bednarek2012,
Weissenborn2012a,Weissenborn2012b,Katayama2012,Colucci2013,Dalen2014,Whittenbury2014,
Katayama2015,Gomes2015,Oertel2015,Maslov2015,Tolos2017,Fortin2017,
Marques2017,Spinella-PhD,Kolomeitsev2017,Raduta2018}, 
delta isobars~\cite{Pandharipande1971,Sawyer1972,Boguta1982,Glendenning1985,Prakash1992,
Chenyj2009,Schurhoff2010,Lavagno2010,Drago2014a,Drago2014b,Caibj2015,Drago2016,Zhuzy2016},
and deconfined quark matter~\cite{Haensel1986,Heiselberg1993,Ghosh1995,Drago2001,
Burgio2002,Weber2005,Alford2005,Ippolito2008,Dexheimer2010,
Bonanno2012,Masuda2013,Orsaria2014,Kojo2016EPJA,Spinella2016,
Fukushima2016,Ranea2017,Sedrakian2017,Alford2017,Alvarez2017,Blaschke2018}.
The details of the composition of compact stars at high densities are
not fully understood yet. The current observational programs focusing
on neutron stars combined with the nuclear physics modeling of
their interiors are aimed at resolving the puzzles associated with
their EoS and interior composition.

Although the appearance of $\Delta$'s in neutron star matter was
conjectured long ago~\cite{Sawyer1972,Pandharipande1971} there has
been much less research on their properties in the intervening years
as compared to hyperons and quark matter. This may partially be a
consequence of Ref.~\cite{Glendenning1985} where $\Delta$'s were found
to appear at densities that are much larger than the typical central
densities of neutron stars. Thus, $\Delta$'s have been considered
largely unimportant in neutron star astrophysics.

Recently, a number of studies of $\Delta$'s in neutron star matter
appeared which were conducted within the CDF theory in the Hartree
approximation, i.e., the so-called relativistic mean-field
model~\cite{Chenyj2009,Schurhoff2010,Lavagno2010,Drago2014a,Drago2014b,
Caibj2015,Zhuzy2016,Drago2016,Spinella-PhD,Kolomeitsev2017}.
Some of these studies ignore hyperons in order to isolate the effects
$\Delta$ isobars have on the nucleonic EoS and neutron star properties
by choosing a particular set (in some cases several sets) of
meson-$\Delta$ coupling constants~\cite{Chenyj2009,Caibj2015,Zhuzy2016,
Kolomeitsev2017}. The universal coupling scheme is typically adopted
in these studies. In analogy to hyperons, the $\Delta$ degrees of 
freedom were found to soften the EoS of neutron star matter and to
reduce the maximum mass of a compact star. However, a simultaneous 
treatment of hyperons and $\Delta$'s appears to be mandatory in order
to assess the overall effect of these new degrees of freedom on dense
matter and the gross properties of compact stars.

{The $\Delta$ degrees of freedom in nuclear dynamics
  have been studied in a number of alternative settings.  $\Delta$'s
  play an important role in the studies of nucleon-pion-$\Delta$
  dynamics, which resum the RPA diagrams including $\Delta$-hole loops
  with the $\Delta$-hole vertex given by $g'_{N\Delta}$ Landau-Migdal
  parameter~\cite{Migdal1990,Herbert1992,Korpa2009}. These
  studies are mainly focused on the pion propagator and dispersion
  (condensation) in nuclear matter.  More recently, $\Delta$'s were
  included in the studies of nuclear matter in the chiral approach
  where the nuclear density functional is arranged in powers of
  small parameters  (e.g. number of derivatives of the pion field) and
  $\Delta$'s appear in virtual states~\cite{Fritsch2005}. }

The principal aim of this work is to explore, in great detail,
the competition between $\Delta$ isobar and hyperon populations in
dense matter, and to study the impact of $\Delta$ populations on the
properties of compact stars such as masses and radii. For that
purpose we carry out a detailed analysis of the parameter space of
the meson-$\Delta$ coupling values within the CDF theory at the
relativistic Hartree and Hartree-Fock level.

This work is organized as follows. In Sec.~\ref{sec:theor_model} we
outline the CDF model and its parametrizations. Section~\ref{sec:results}
presents our results for the EoS of dense matter and its composition. 
The global properties of compact stars and their internal structures are
discussed in this section as well. Finally, a summary of our results is 
provided in Sec.~\ref{sec:conclusions}.

%-------
\section{Theoretical model}
\label{sec:theor_model}
%-------

%---------
\subsection{CDF model for stellar matter}
\label{sec:CDF}
%--------

We start with a brief outline of our theoretical framework, which is
based on the CDF theory treated in the Hartree and Hartree-Fock
approximations. The Lagrangian density of the model is given by
\begin{equation}
\mathscr{L} = \mathscr{L}_B + \mathscr{L}_{m} + \mathscr{L}_{\text{int}} + \mathscr{L}_l,
\end{equation}
where the first term $\mathscr{L}_B$ is the Lagrangian of free
baryonic fields $\psi_B$, with index $B$ labeling the spin-1/2
baryonic octet, which comprises nucleons $N \in \{n, p\}$,
hyperons $Y \in \{ \Lambda, \Xi^{0,-}, \Sigma^{+,0,-} \}$, and the
spin-3/2 zero-strangeness quartet $\Delta \in \{ \Delta^{++, +, 0,-} \}$. 
Note that the $\Delta$'s are treated as Rarita-Schwinger 
particles~\cite{Pascalutsa2007}. The second term $\mathscr{L}_{m}$ 
represents the Lagrangian of free meson fields $\phi_m$, which are 
labeled according to their parity, spin, isospin and strangeness. 
In the present model we include the isoscalar-scalar meson $\sigma$, 
which mediates the medium-range attraction between baryons, the isoscalar-vector
meson $\omega$, which describes the short range repulsion, the 
isovector-vector meson $\rho$, which accounts for the isospin 
dependence of baryon-baryon interactions, and the $\pi$ meson which
accounts for the long-range baryon-baryon interaction and the tensor force. 
The two hidden-strangeness mesons, $\sigma^\ast$ and $\phi$, describe
interactions between hyperons. The interaction between the baryons
and mesons is described by the third term $\mathscr{L}_{\text{int}}$ which
has the generic form
\begin{equation}
\label{eq:Lagrangian}
\mathscr{L}_{\text{int}} \equiv g_{mB}\tau_B\bar{\psi}_B\Gamma_m\varphi_m\psi_B,
\end{equation}
where $g_{mB}$ is the meson-baryon coupling constant,
$\tau_B \in \{1, \bm{\tau}\}$ is the isospin matrix and
$\Gamma_m \in \{1,\gamma_\mu,\gamma_5\gamma_\mu,\sigma_{\mu\nu}\}$ is
the relevant (Dirac-matrix) vertex. Finally, the last term $\mathscr{L}_l$
describes the contribution from free leptons; we include electrons ($e^-$) and
muons ($\mu^-$) and neglect the neutrinos which are irrelevant at low temperatures.

Starting from Eq.~\eqref{eq:Lagrangian} we carry out the standard
procedure for obtaining the density functional in CDF theories. This
amounts to finding the equations of motions from the Euler-Lagrange
equations of the theory, which for the baryon octet and leptons
have the form of the Dirac equation, whereas for the $\Delta$
decuplet are given by the Rarita-Schwinger equation. The equations 
of motion for meson in the mean-field approximation take the form of
Klein-Gordon equation. Each of the baryon self-energies is then
decomposed in the Dirac space according to
\begin{equation}
\Sigma(k) = \Sigma_S (k) + \gamma_0 \Sigma_0 (k) + \bm{\gamma}\cdot \hat{\bm{k}}\Sigma_V (k)
\end{equation}
%-
where $\Sigma_S$, $\Sigma_0$ and $\Sigma_V$ are the scalar, time and
space components of the vector self-energies and $\hat{\bm{k}}$ is a
unit vector along $\bm{k}$. The energy density functional is then 
generated by evaluating the baryon self-energies  $\Sigma(k)$ in the 
Hartree (RMF) or Hartree-Fock (RHF) approximations~\cite{Bouyssy1987,
Ring1996,Serot1997,Long2006}. The detailed expressions for self-energies
are given, for instance, in Refs.~\cite{Zhuzy2016,Lijj2018}. Note that
the pion-exchange and the tensor couplings of vector mesons to baryons
contribute only to the Fock self-energies. In $\beta$-equilibrium the
chemical potentials of the particles are related to each other by
\begin{equation}
\mu_B = b_B \mu_n - q_B \mu_e,
\end{equation}
where $b_B$ and $q_B$ denote the baryon number and electric charge of
baryon species $B$, and $\mu_n$ and $\mu_e$ are the chemical
potentials of neutrons and electrons, respectively. This, together
with the field equations and charge neutrality condition allows
us to determine the EoS and composition of matter for any given net
baryonic density $\rho$ at zero temperature self-consistently.

Once the EoS is determined, the integral parameters, in particular the
mass and the radius, of a compact star of given central density can be
computed from the Tolman-Oppenheimer-Volkoff (TOV)
equations~\cite{Tolman1939,Oppenheimer1939}. To do so we match
smoothly our EoS to an EoS of the inner and outer
crusts~\cite{Baym1971a,Baym1971b} at the crust-core transition density
$\rho_0/2$, where $\rho_0$ denotes the saturation density of ordinary 
nuclear matter.

%----------------
\subsection{Meson-baryon couplings}
\label{sec:couplings}
%----------------

We now turn to the procedure of choosing the appropriate values of the
coupling constants $g_{m\Delta}$ between the mesons and baryons. These
have to be fitted to the experimental (empirical) data of nuclear and
hypernuclear systems.  {In the purely nucleonic sector the
meson-nucleon ($mN$) couplings are given by
%-----------------------------------------------------------------------
$
g_{m N}(\rho)=g_{m N}(\rho_0)f_{m N}(x),
$
%-----------------------------------------------------------------------
where $x=\rho/\rho_0$, $\rho$ is the baryonic density.
For the isoscalar channel, one has
%-----------------------------------------------------------------------
\begin{align}
f_{m N}(x)=a_m\frac{1+b_m(x+d_m)^2}{1+c_m(x+d_m)^2}, \quad m = \sigma, \omega,
\end{align}
%-----------------------------------------------------------------------
which is subject to constraints $f_{m N}(1)=1$,
$f^{\prime\prime}_{m N}(0)=0$ and
$f^{\prime\prime}_{\sigma N}(1)=f^{\prime\prime}_{\omega N}(1)$. 
The density dependence for the isovector channels is taken in an
exponential 
form~\footnote{For the PKO3 interaction used in this study the masses
  (in MeV) of nucleon, $\sigma$-, $\omega$-, $\rho$- and $\pi$-mesons
  are $938.9$, 525.6677, $783$, 769, 138.  The coupling constants at
  the saturation $\rho_0 =0.153$ fm$^{-3}$ are $g_\sigma = 8.8956$,
  $g_\omega = 10.8027$, $g_\rho = 2.0302$ and $f_\pi = 0.3929$; the
  remaining parameters, which describe the density-dependence of
  couplings can be found in Table 1 of Ref.~\cite{Long2008}.}} 
%-----------------------------------------------------------------------
\begin{align}
f_{mN}(x) = e^{-a_m (x-1)}, \quad m = \rho, \pi.
\end{align}
%-----------------------------------------------------------------------
In the
hypernuclear sector, as usual, the vector meson-hyperon couplings are
given by the SU(3) flavor symmetric quark
model~\cite{Swart1963,Schaffner1994} whereas the scalar meson-hyperon
couplings are determined by their fitting to empirical hypernuclear
potentials.  {We note that the isovector couplings are
  non-universal and, for example, values
  $g_{\rho\Sigma}/g_{\rho N} \simeq 1/4$-$1/3$ are required to
  describe the $\Sigma$-atom~\cite{Mares1995}.  }

\begin{table}[tb]
  \caption{ The parameters of symmetric nuclear matter at saturation
    density and the masses and radii of hypernuclear stars, predicted
    by the hypernuclear CDF theory with PKO3 and DD-ME2
    parametrizations. Upper panel: the saturation density $\rho_0$
    (fm$^{-3}$), binding energy $E_{B}$ (MeV), compression modulus $K$
    (MeV), symmetry energy $J$ (MeV) and its slope $L$ (MeV), the
    Dirac mass $M^\ast_D$ (in units of nucleon mass $M$), and the
    Landau mass $M^\ast_{L}$ ($M$) for symmetric
    nuclear matter. Lower panel: The  mass $M_{\text{max}}$ (in solar units),
    radius $R_{\text{max}}$ (km) and central density $\rho_{\text{max}}$ (fm$^{-3}$)
    of  {the maximum-mass} star, the threshold density $\rho^{\text{c}}_Y$ (fm$^{-3}$) for
    hyperons $\Lambda$ and $\Xi^-$, the radius $R_{1.4}$ (km) and
    central density $\rho_{1.4}$ (fm$^{-3}$) for  a canonical
    $1.4M_\odot$ neutron star.} \setlength{\tabcolsep}{3.0pt}
\label{tab:MP}
\begin{tabular}{cccccccccccc}
\hline\hline
\multirow{2}*{CDF} & \multicolumn{7}{c}{Symmetric nuclear matter} \\
\cline{2-8}    & $\rho_0$  & $E_{B}$   & $K$     & $J$    & $L$     & $M^\ast_D$ & $M^\ast_{L}$ \\
\hline
PKO3           & 0.153     & $-16.04$  &  262.44 &  32.99 &  82.99  & 0.59       & 0.72 \\
DD-ME2         & 0.152     & $-16.14$  &  251.15 &  32.31 &  51.27  & 0.57       & 0.63 \\
\hline
\multirow{2}*{ } & \multicolumn{7}{c}{Hypernuclear matter} \\
\cline{2-8}    & $M_{\text{max}}$ & $R_{\text{max}}$ & $\rho_{\text{max}}$ & $\rho^{\text{c}}_\Lambda$ & $\rho^{\text{c}}_\Xi$ & $R_{1.4}$ & $\rho_{1.4}$ \\
\hline
PKO3           & 2.00      & 11.82     &  0.96        &  0.33       &  0.48   & 13.96     & 0.31 \\
DD-ME2         & 2.00      & 11.83     &  0.93        &  0.34       &  0.38   & 13.22     & 0.34 \\
\hline\hline
\end{tabular}
\end{table}

Let us focus now on the range of the meson-$\Delta$ couplings. No consensus
has been reached yet on the magnitude of the $\Delta$ potential in nuclear 
matter. The phenomenological model analyses of the scattering of electrons 
and pions off nuclei and photoabsorption~\cite{Wehrberger1989,Connell1990,
Alberico1994,Nakamura2010} indicate that the $\Delta$ isoscalar potential 
$V_\Delta$ should be in the range~\cite{Drago2014a}
\begin{equation}
\label{eq:scattering_constraint}
-30~\text{MeV} + V_N(\rho_0) \lesssim V_\Delta(\rho_0) \lesssim V_N(\rho_0),
\end{equation}
where $V_N = \Sigma_{0,\omega(\sigma)} + \Sigma_{S,\sigma(\omega)}$ is the
nucleon isoscalar potential. The studies of $\Delta$ production in heavy-ion
collisions~\cite{Ferini2015,Guowm2015,Cozma2016} suggest a less attractive 
potential~\cite{Kolomeitsev2017},
\begin{equation}
\label{eq:HIC_constraint}
V_N(\rho_0) \lesssim V_\Delta(\rho_0) \lesssim 2/3V_N(\rho_0).
\end{equation}
Below, we will use instead of $g_{m \Delta}$ the ratio $ R_{m \Delta}
= g_{m \Delta}/ g_{m N}. $ The isoscalar potential of the $\Delta$
isobars in symmetric nuclear matter at saturation density is thus given by
\begin{equation}
V_\Delta(\rho_0) =  R_{\omega\Delta}\Sigma_{0(S),\omega}(\rho_0) + R_{\sigma\Delta}\Sigma_{S(0),\sigma}(\rho_0).
\end{equation}
The isovector meson-$\Delta$ couplings are largely unknown.  It has
been found that the critical density of the onset of $\Delta$'s is most
sensitive to the ratio
$R_{\rho\Delta}$~\cite{Drago2014a,Caibj2015}. In our numerical study
we use two representative parametrizations of the nucleonic CDF theory
based on the density-dependent meson-baryon couplings, specifically
the relativistic Hartree-Fock PKO3 parametrization~\cite{Long2008} and
the relativistic Hartree DD-ME2
parametrization~\cite{Lalazissis2005}. Both parametrizations describe
successfully the properties of finite nuclei. We will use the
extensions of these models to the hypernuclear sector as given in
Ref.~\cite{Lijj2018}. In this work the meson-hyperon couplings have
been chosen to: (a) reproduce the empirical potentials of hyperons in
nuclear matter deduced from nuclear structure calculation and (b)
produce heavy $2M_\odot$ compact stars~\cite{Antoniadis2013}. In
Table~\ref{tab:MP} we list the key parameters of symmetric nuclear
matter and some properties of hyperonic stars predicted by the two
models. Note that the combined analysis of terrestrial experiments and
astrophysical observations~\cite{Oertel2017} predict values for the
symmetry energy $J = 31.7 \pm 3.2$~MeV and its slope
$L = 58.7\pm 28.1$~MeV at saturation density. The $J$ value predicted
by both parametrizations are compatible with the central value of
32~MeV, while the value of $L$ predicted by PKO3 is located at the
upper bound of the preferred range.

%-------------
\section{Results and discussions}
\label{sec:results}
%-------------

In this section we investigate the competition between $\Delta$
isobars and hyperons in dense stellar matter and their effect on the
properties of compact stars. In a first step, we will analyze the EoS
of stellar matter with $\Delta$'s for selected sets of $R_{m\Delta}$
values, which are motivated by the constraints shown in Eqs.
~\eqref{eq:scattering_constraint} and~\eqref{eq:HIC_constraint}; note
that the value of the potential $V_\Delta$ does not fix any of the
couplings, rather it provides a relation between $R_{\omega\Delta}$
and $R_{\sigma\Delta}$. This illustrates the general features that
emerge when $\Delta$ isobars are included. In the second step, we
construct the EoS and the associated stellar models for a continuum of
meson-$\Delta$ couplings which span the complete parameter space. 
This provides insight into the dependence of the stellar
parameters on $R_{m\Delta}$, which, in turn, permits us to
narrow down the meson-$\Delta$ parameter space using the
astrophysical constraints.

%---------------------------------------------------------------
\subsection{Illustrative cases}
\label{sec:illustration}
%---------------------------------------------------------------
\begin{figure}[tb]
\centering
%\ifpdf
\includegraphics[width = 0.44\textwidth]{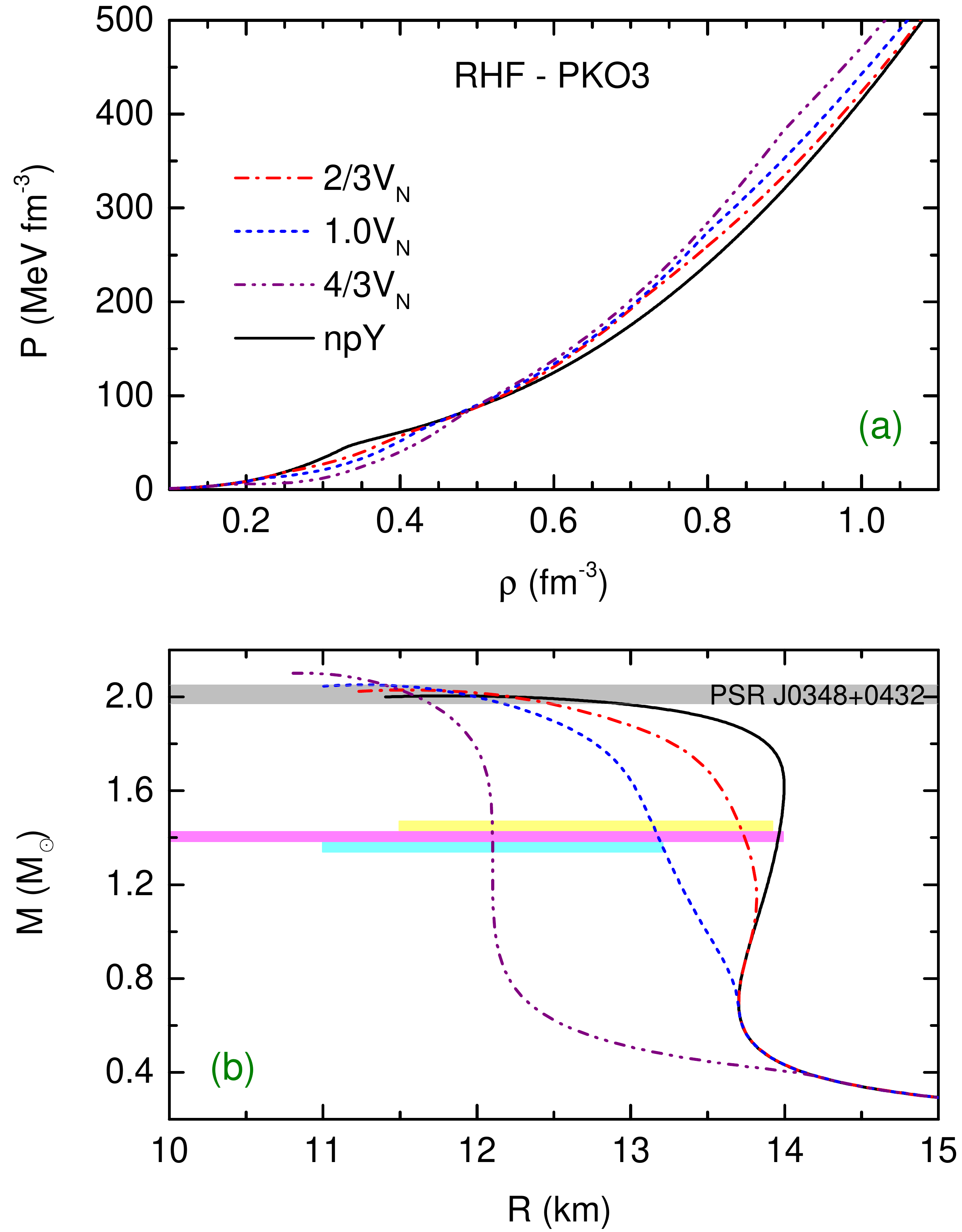}
%\else
%\includegraphics[width = 0.44\textwidth]{EoS.eps}
%\fi
\caption{Effects of $\Delta$-isobars on the EoS and mass-radius
  relation of compact stars. (a) EoSs for different $\Delta$
  potential depths $V_\Delta(\rho_0) = (1\pm1/3) V_N (\rho_0)$, $npY$
  denotes the hyperonic matter, (b) the corresponding mass-radius
  relations.  The gray shading indicates the mass of PSR
  J0348+0432. The yellow, pink, and cyan shading indicate the radius
  range of canonical 1.4$M_\odot$ neutron stars set by
  Refs.~\cite{Hebeler2013,Lattimer2014, Steiner2017}. Notice that the
  ordering of the EoSs according to their stiffness depends on the
  density interval. The $\omega$-$\Delta$ coupling constants are fixed
  as 1.1, while the $\sigma$-$\Delta$ coupling constants are adjusted
  to the potential depths $V_\Delta(\rho_0)$,  {which corresponds to
  $R_{\sigma\Delta} = 1.075 \pm 0.125$.} The results are calculated for
  the RHF parametrization PKO3.}
\label{fig:EoS}
\end{figure}
%---------------------------------------------------------------

We start with several illustrative examples which are constructed as
follows: (a) the $R_{\omega\Delta}$ parameter is kept fixed at a
value of $1.1$, which leads to the largest masses of compact star 
for our parameter space (to be discussed in Fig.~\ref{fig:RHF} below); 
(b) all isovector meson couplings are set equal to the meson-nucleon 
coupling; (c) the potential $V_\Delta$ is varied  within the bounds 
set by Eqs.~\eqref{eq:scattering_constraint} and \eqref{eq:HIC_constraint}
by tuning the $R_{\sigma\Delta}$ parameter.

In Fig.~ {\ref{fig:EoS}(a)} we show the EoS of $\Delta$-admixed-hypernuclear
($npY\Delta$) matter for three values of the isoscalar potential 
$V_\Delta(\rho_0)$. For comparison, we also show the case of matter 
without $\Delta$'s, i.e., $npY$ matter. The corresponding mass-radius
relations of compact-star models computed for these EoS are shown 
in Fig.~\ref{fig:EoS}(b). In the case of $npY$ matter the abrupt change
in the slope of the pressure at baryonic density $\rho \simeq 0.33$~fm$^{-3}$
is the result of the onset of hyperons in matter. It is seen that if 
$\Delta$'s are included in the composition, the EoS is softened at 
low and stiffened at high densities. This is the more pronounced the
deeper the $V_\Delta(\rho_0)$ potential.

These modifications in the EoS affect the mass-radius relations of
neutron-star models as depicted in Fig.~\ref{fig:EoS}(b). It is
seen that accounting for $\Delta$'s reduces the radii of models from
their values obtained for hypernuclear and nuclear matter EoS. This
is a consequence of the {\it softening of the EoS} at low to
intermediate baryonic densities.  Note that the $\Delta$'s may
appear already at densities slightly above the nuclear saturation
density, which implies that even low-mass ($\simeq M_{\odot}$)
compact stars can be affected by $\Delta$ populations. As
expected, the larger the $V_\Delta(\rho_0)$ potential, the larger
is the observed shift in the radius. For example, for
$V_\Delta(\rho_0) = 4/3V_N(\rho_0)$ the radius of a canonical 
$1.4M_\odot$ neutron star is about 2~km smaller than the
radius of its purely hyperonic or nucleonic counterpart. At the
same time, $\Delta$'s lead to marginally greater maximum masses of
compact stars, because of the {\it stiffening of the EoS} in the
high-density region. The deeper the potential $V_\Delta(\rho_0)$
the larger the star's maximum mass.

%---------------------------------------------------------------
\begin{figure}[tb]
\centering
%\ifpdf
\includegraphics[width = 0.44\textwidth]{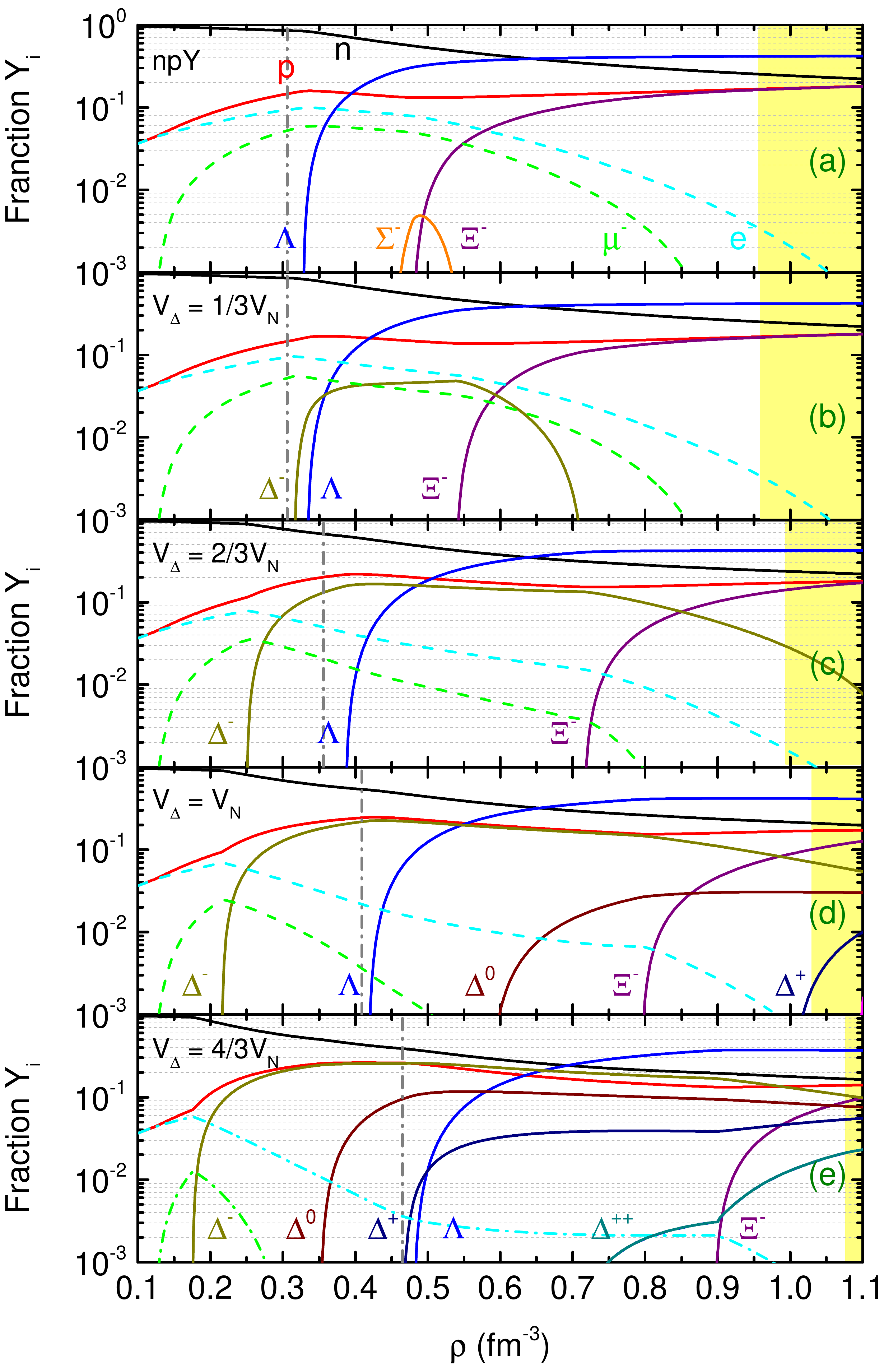}
%\else
%\includegraphics[width = 0.44\textwidth]{Frac.eps}
%\fi
\caption{Particle fractions {in $npY$ (panel a) and $npY\Delta$
  (panels b, c, d and e) matter. The $\Delta$ potential depths are}
  $V_\Delta = 1/3V_N$ (b), $2/3V_N$ (c), $V_N$ (d) and $4/3V_N$ (e).
  The thick vertical lines indicate the central density of the
  respective canonical $1.4M_\odot$ neutron star, the yellow shadings
  indicate densities beyond the maximum mass configurations.  The
  results are calculated by using the RHF parametrization PKO3.}
\label{fig:Frac}
\end{figure}
%---------------------------------------------------------------

Figure~\ref{fig:Frac} shows the particle fractions for several
selected values of  {the} $V_\Delta(\rho_0)$ potential. In the
case of hyperonic matter without $\Delta$'s, shown in 
Fig.~\ref{fig:Frac}(a), the first hyperon to appear is the $\Lambda$,
which is followed by the $\Xi^-$ hyperon. The $\Sigma^-$ hyperons
appear only briefly, because they are disfavored due to their
repulsive potential at nuclear saturation density. This sequence of
hyperon thresholds is consistent with the recent hypernuclear CDF
computations of Refs.~\cite{Weissenborn2012b,Fortin2017,Spinella-PhD,
Raduta2018}. In the cases when $\Delta$'s are taken into account 
the following new features are observed: (1) The $\Delta$-threshold 
density could be at much lower density than that for the first hyperon
(i.e., the $\Lambda$). The larger the potential $V^{(N)}_\Delta$ the 
lower the onset threshold for $\Delta$'s. (2) The first $\Delta$ 
isobar to appear is the $\Delta^-$, which eliminates the $\Sigma^-$
entirely and significantly shifts the threshold for the
$\Xi^-$; the threshold of the $\Lambda$ hyperon is also shifted
but to a lesser extent. (3) In the case of a strongly
attractive potential $V_\Delta \leq V_N$, the $\Delta^{0,+}$
resonances appear at intermediate to high densities, in addition to 
the $\Delta^-$. (4) Electric charge neutrality, maintained by baryons
and leptons, implies that the onset of the $\Delta^-$ not only shifts
the threshold of $\Xi^-$ hyperons, but leads also to a depletion of 
the negatively charged lepton population, especially that of muons.
(5) For $V_\Delta \geq 2/3V_N$ the threshold for the onset of 
$\Delta$'s is reached in a canonical $1.4M_{\odot}$ neutron star. 
Furthermore, the central density of the maximum-mass star is larger
by about 0.1~fm$^{-3}$ than in the absence of $\Delta$'s.

\begin{table}[tb]
  \caption{ The threshold densities {(in fm$^{-3}$)} for the
    onset of direct Urca processes in $npY$ matter (rows 1 and 3) and
    in $npY\Delta$ matter with $V_\Delta(\rho_0) = V_N(\rho_0)$ (rows
    2 and 4). No entry means that the process is forbidden.  The rows
    1 {and} 2 correspond to PKO3, and {rows 3 and 4} to {the}
    DD-ME2 parametrizations. The thresholds (from left to right)
    corresponds to the following processes: $n\to p + e^- + \bar
    \nu_e$, $\Delta^-\to \Lambda + e^- + \bar \nu_e$, $\Lambda\to p +
    e^- + \bar \nu_e$, and $\Xi^-\to \Lambda + e^- + \bar \nu_e$.  }
  \setlength{\tabcolsep}{6pt}
\label{tab:DU}
\begin{tabular}{cccccc}
\hline\hline
 CDF & Composition & $\rho^{\text{DU}}_n$ & $\rho^{\text{DU}}_{\Delta^-}$ & $\rho^{\text{DU}}_\Lambda$ & $\rho^{\text{DU}}_{\Xi^-}$ \\
\hline
\multirow{2}*{PKO3}    & $npY$       &  0.282     &  -                  &  0.329           &  0.536           \\
                       & $npY\Delta$ &  0.271     &  0.431              &  0.458           &  -               \\
\\
\multirow{2}*{DD-ME2}  & $npY$       &  -         &  -                  &  0.341           &  0.382           \\
                       & $npY\Delta$ &  -         &  0.358              &  0.366           &  -               \\
\hline\hline
\end{tabular}
\end{table}

The inclusion of the $\Delta$-isobars in the EoS of neutron stars can
impact their cooling through modifications of the direct Urca (DU)
neutrino emissivities in the star's core. The onsets densities of
the electronic versions of DU processes in purely $npY$ and
$npY\Delta$ matter are listed in Table~\ref{tab:DU}. We recall
that for $npY$ matter, the nucleonic DU process
($n \rightarrow p + e^- + \bar{\nu}_e$) occurs only for the PKO3
parametrization, while the hyperonic DU processes operate for
both PKO3 and DD-ME2 parametrizations as soon as hyperon states are
populated~\cite{Lijj2018}. In $npY\Delta$ matter the early appearance
of the $\Delta^-$ causes a significant increase in the proton ($p$)
fraction but a decrease in the $e^-$ fraction (see Fig.~\ref{fig:Frac}).
As a result, the onset of the nucleonic DU process is slightly shifted
toward lower densities for the PKO3 parametrization. This also results
in a delay in the onset of the $\Lambda\rightarrow p + e^- + \bar{\nu}_e$
process as the density increases. The $\Delta^- \rightarrow \Lambda + 
e^- + \bar{\nu}_e$ process proceeds for both parametrizations shortly
after the $\Lambda$'s appear, which occurs for stars with masses
$M \ge 1.5M_\odot$. The $\Xi^- \rightarrow \Lambda + e^- + \bar{\nu}_e$ 
process is absent at high density due to the very low fraction of
leptons. Our speculations above show that the inclusion of
$\Delta$'s may cause substantial modifications in the rates at which
neutrinos are emitted from neutron stars. It will be worthwhile
to explore these modifications in numerical cooling simulations.

%------
\subsection{Meson-$\Delta$ coupling space}
\label{sec:m_delta_coupling}
%------

%---------------------------------------------------------------
\begin{figure}[tb]
\centering
%\ifpdf
\includegraphics[width = 0.48\textwidth]{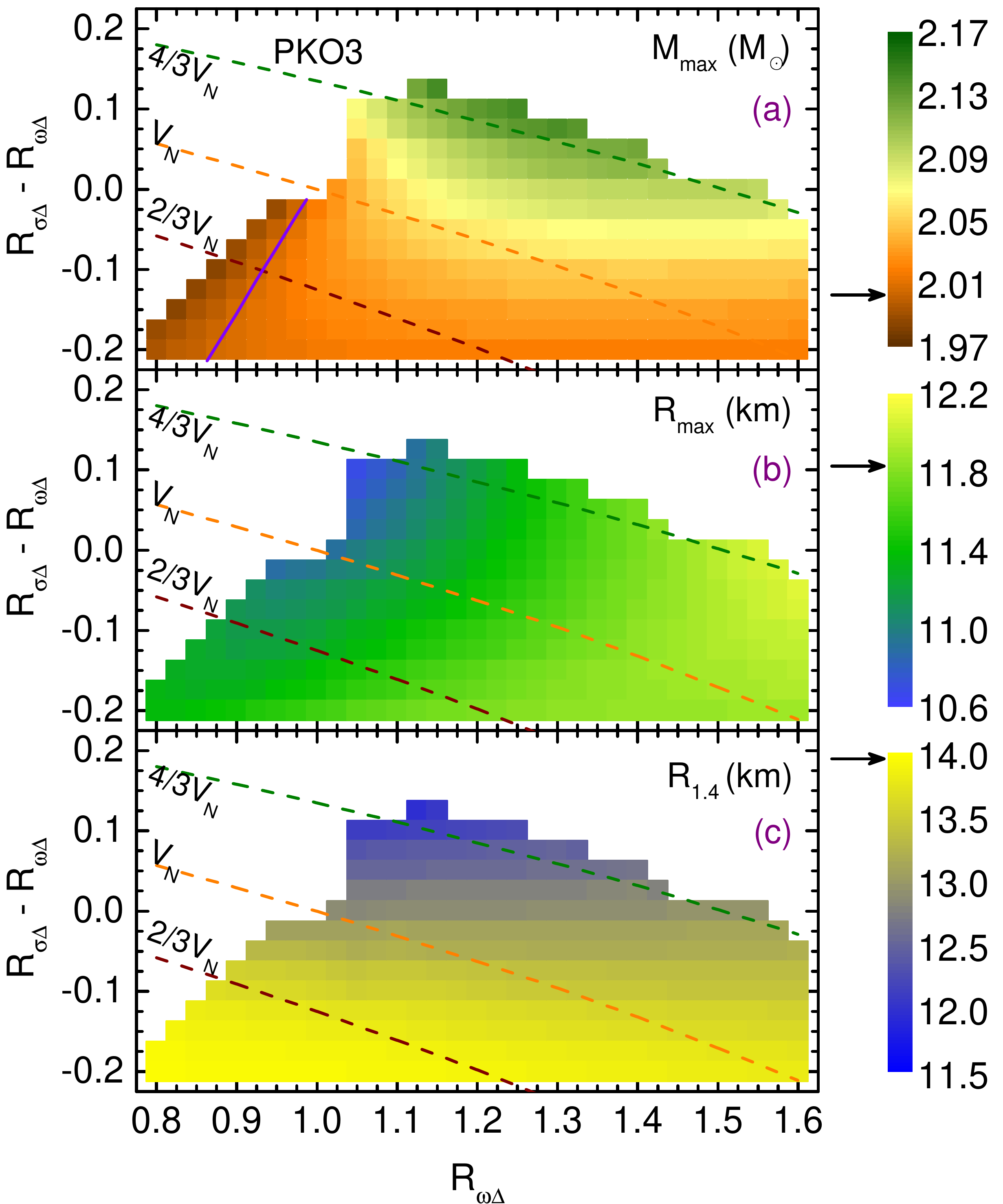}
%\else
%\includegraphics[width = 0.48\textwidth]{RHF.eps}
%\fi
\caption{ Contour plots for the gross properties of compact stars in
  the parameter space spanned by $R_{\omega\Delta}$ and
  $(R_{\sigma\Delta}-R_{\omega\Delta})$. Shown are the: (a) maximum
  mass, (b) radius of the maximum mass star, and (c) radius of a
  canonical $1.4M_\odot$ star. The dashed lines show the constant
  values of the potential $V_\Delta = (1\pm1/3)V_N$. The solid line
  in the lower-left corner of panel (a) means that the configurations
  contain $\Delta$'s but have a mass equal to the purely hyperonic
  star. The white areas indicate coupling sets for which no physical
  solutions exist. The mass and radii for purely hyperonic matter
  are marked by horizontal arrows pointing at the labels. To obtain
  these results we used the RHF parametrization PKO3.
   {Analogous contour plots for relativistic mean-field
 models were first obtained in Ref.~\cite{Spinella-PhD}}. 
}
\label{fig:RHF}
\end{figure}
%---------------------------------------------------------------
%
%---------------------------------------------------------------
\begin{figure}[tb]
\centering
%\ifpdf
\includegraphics[width = 0.48\textwidth]{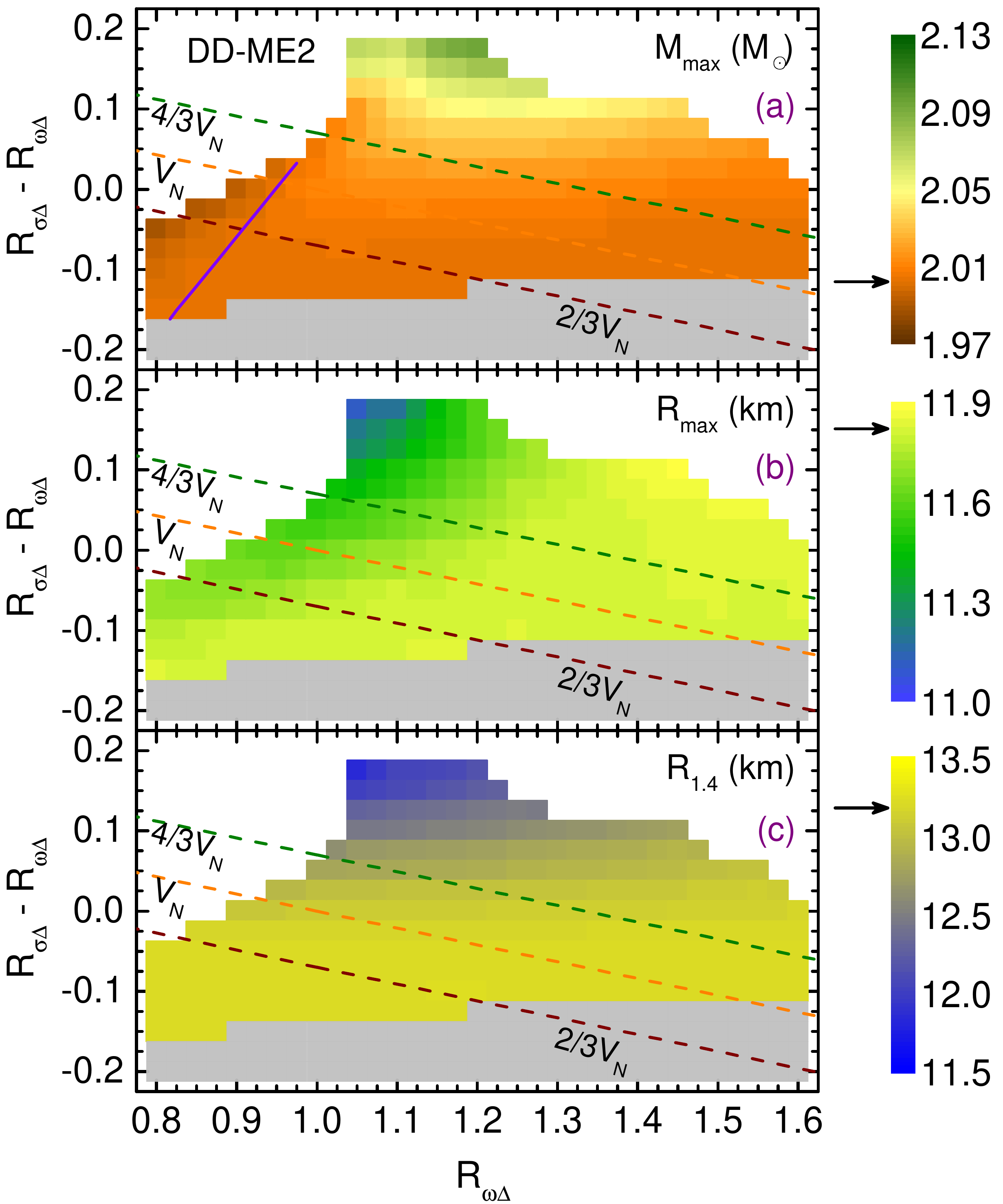}
%\else
%\includegraphics[width = 0.48\textwidth]{RMF.eps}
%\fi
\caption{Same as Fig.~\ref{fig:RHF}, but for the RMF parametrization DD-ME2.
The grey pixels show areas where no $\Delta$-isobars are populated.}
\label{fig:RMF}
\end{figure}
%---------------------------------------------------------------

Having established some general trends, we would now like to explore
the parameter space in a more systematic manner. We will still keep
the isovector meson-$\Delta$ couplings in accord with the universal
scheme, i.e., $R_{\rho(\pi)\Delta} = 1.0$. Instead of having
$R_{\omega\Delta} = 1.1$ we will now allow for variations of
$R_{\omega\Delta} \in [0.8;1.6]$ and
$R_{\sigma\Delta}-R_{\omega\Delta} \in [-0.20; 0.20]$. As we will 
see below, such a range captures the most interesting region of the
parameter space spanned by the masses and radii of the models.

The three panels in Fig.~\ref{fig:RHF} show the value of the
maximum-mass star, the radius of this star, and the radius of a
canonical $1.4M_{\odot}$ star, computed for the PKO3 parametrization.
Figure~\ref{fig:RMF} shows the same quantities but computed for
the DD-ME2 parametrization. The white areas indicate the range of
couplings for which the EoS is unphysical due to either a negative
Dirac baryon mass or a non-monotonic pressure. The grey pixels
indicate that no $\Delta$-isobars are populated. The lines of
constant values of the potential $V_\Delta = (1\pm1/3)V_N$ are also
shown.

Figures~\ref{fig:RHF} and~\ref{fig:RMF} display some common features
on which we comment first; we will return to differences below. It is
seen that (a) the maximum masses increase when the couplings
$R_{m\Delta}$ are increased, i.e., when moving from the lower-left
corner to the right. For $R_{m\Delta}\le 1.0$, i.e., when $\Delta$'s
interact weaker than nucleons, the stellar masses are close or
slightly below the $2M_{\odot}$ limit (i.e., the purely hyperonic
case). For the parameter space considered here, the stellar
masses are consistent with the current observational limits on pulsar
masses. The heaviest stars appear when both $V_\Delta$ is most
attractive and the difference $(R_{\sigma\Delta}-R_{\omega\Delta})$ is
largest. (b) The smallest radii of the maximum-mass stars are located in the
region where $R_{\omega\Delta }\sim 1.0$ and $R_{\sigma\Delta}\sim 1.1$, i.e.,
the most massive compact stars are not automatically also the most
compact ones. Quite generally, the radii of stars containing 
$\Delta$'s in their cores are smaller than the radii of their
nucleonic/hyperonic counterparts. (c) For neutron stars with 
canonical masses of around $1.4 M_\odot$ we find a strong reduction
of the radius of about $\le 2.5$~km. The most compact stars, in
this case, are those {for which the coupling $R_{\sigma\Delta}$ is
maximal. This reduction {may help to achieve a better agreement of 
the theoretical model parameters of neutron stars with observations. 
Indeed the models with hyperons only produce $R=13.9$~km for a 
$M=1.4M_{\odot}$ model star, which is close to the upper range of
radii (14~km) inferred in Refs.~\cite{Hebeler2013,Steiner2017,Fattoyev2017}.
But it fails to satisfy the 13.2~km upper bound obtained in
Refs.~\cite{Lattimer2013,Lattimer2014}. However, if $\Delta$'s are
included and the coupling constants are chosen such that
$R_{\sigma\Delta} \geq R_{\omega\Delta}$, the radius of the
$M=1.4M_{\odot}$ model is sufficiently reduced so that the latter 
constrained is satisfied too.

The trends discussed above can also be seen in Fig.~\ref{fig:RMF}.
One sees that for the PKO3 parametrization $\Delta$'s appear already
for $R_{\omega\Delta}-R_{\sigma\Delta} = -0.20$ irrespective of the
value of $R_{\omega\Delta}$, while for the DD-ME2
parametrization $\Delta$'s are absent when
$ R_{\omega\Delta}-R_{\sigma\Delta} \leq -0.10$. This can be traced
back to the differences in the isoscalar and isovector sectors of the
{\it nucleonic CDF} models~\cite{Caibj2015}, as it is the nucleonic
CDF model that determines the critical density of $\Delta$.

We have also examined the dependence of the compact star properties on
the isovector meson-$\Delta$ coupling by setting $V_\Delta = V_N$ and
$R_{\omega\Delta} = 1.1$ while varying $R_{\rho(\pi)\Delta}$ in the
range $[0; 3]$. (We recall that the $V_\Delta(\rho_0)$-potential is
independent of $R_{\rho(\pi)\Delta}$.)   {We find that
  modifications of the isovector couplings change the critical
  densities of delta isobars (for example, an increase $\simeq 0.04$
  fm$^{-3}$ for $\Delta^-$ is observed over the entire range), which
  changes the particle fractions. Nevertheless,} the maximum-star mass
is insensitive to changes in $R_{\rho(\pi)\Delta}$. Indeed, varying
$R_{\rho(\pi)\Delta}$ within the bounds mentioned just above decreases
the maximum mass by only about} 0.02$M_\odot$ for both
parametrizations. The radii are almost unchanged for the maximum-mass
configurations, while they increase by about 0.3~km for canonical
1.4$M_\odot$ neutron stars.   {This is because (a) the
  energy and pressure densities are dominated by the isoscalar
  channels and (b) the isovector couplings vanish exponentially at
  large densities in the present model.}  We conclude that there are
no significant changes in stellar parameters associated with the
variations of the isovector meson-$\Delta$ coupling.

%-----------------
\section{Summary and conclusions}
\label{sec:conclusions}
%-----------------

In this work, we have studied models of stellar matter which contain
the full baryon octet and the delta decuplet within the CDF theory. 
We have used a class of models of CDF which feature density-dependent
meson-baryon couplings. The meson-$\Delta$ coupling constants cannot
be unambiguously constrained with the available empirical data.
Therefore we have conducted a parameter study which included
assumptions about the strength of the $\Delta$ potential in nuclear
matter and a variation of couplings of the $\Delta$'s to mesons with
strength close to that for nucleons. Our results indicate that
$\Delta$ isobars may indeed appear in dense nuclear matter for a wide
range of meson-$\Delta$ coupling constants.

We find that the appearance of $\Delta$'s softens the EoS in the low
to intermediate-density region and stiffens it at high densities. 
This has two {important} effects on the global parameters of neutron
stars: firstly, the maximum mass of a compact star increases by a
 {small} amount.  Because the hypernuclear CDF parametrizations,
employed in this work, satisfy the 2$M_\odot$ maximum mass constraint,
the inclusion of $\Delta$'s affect this feature only mildly.
Secondly, and more importantly, the radius of a compact star
decreases considerably, by about $2$~km, for stars with a canonical
mass of around $1.4M_\odot$ once $\Delta$'s are included.

We argued above that $\Delta$'s may also have a significant effect on
the thermal evolution of compact stars, because they lead to changes
in the particle concentration, which changes the thresholds and
efficiency of the nucleonic and hyperonic DU processes.  Furthermore,
$\Delta^-$'s would contribute to the emissivity of the star via the DU
process $\Delta^- \to \Lambda +e^-+\bar\nu_e$, which has an efficiency
comparable to that of the nucleonic
counterparts~\cite{Prakash1992}. This process requires coexistence of
$\Delta^-$'s and $\Lambda$'s which occurs in stellar models with
$ M \ge 1.5~M_\odot$.  Therefore, a natural extension of the present
work would be to simulate the cooling of neutron stars containing
$\Delta$'s in their centers.

\section*{Acknowledgements}
J. L. acknowledges the support by the Alexander von Humboldt
foundation. A. S. is supported by the Deutsche
Forschungsgemeinschaft (Grant No. SE 1836/4-1). The support from
European COST Actions "NewCompStar" (MP1304) and "PHAROS"
(CA16214) and the LOEWE-Program of Helmholtz International Center for
FAIR of the state of Hesse (Germany) is gratefully acknowledged.
F. W. is supported by the National Science Foundation (USA) under
Grants PHY-1411708 and PHY-1714068.

\end{document}